\documentclass[sort&compress]{elsarticle}

\usepackage{graphicx}
\usepackage{amssymb}
\usepackage{color}
\usepackage{url}
\usepackage{ulem}
\usepackage[figuresright]{rotating}

\journal{Nuclear Instruments and Methods A}

\begin{document}

\begin{frontmatter}


\title{Tests of the Monte Carlo Simulation of the Photon-Tagger Focal-Plane 
Electronics at the MAX IV Laboratory}


\author[lund]{M.F.~Preston}
\author[duke]{L.S.~Myers\fnref{fn1}}
\author[glasgow]{J.R.M.~Annand}
\author[lund]{K.G.~Fissum\corref{cor1}}
\ead{kevin.fissum@nuclear.lu.se}
\author[m4]{K.~Hansen}
\author[m4]{L.~Isaksson}
\author[arktis]{R.~Jebali\fnref{fn2}}
\author[m4]{M.~Lundin}

\address[lund]{Lund University, SE-221 00 Lund, Sweden}
\address[duke]{Duke University, Durham NC 27708, USA}
\address[glasgow]{University of Glasgow, Glasgow G12 8QQ, Scotland, UK}
\address[arktis]{Arktis Radiation Detectors Limited, 8045 Z\"{u}rich, Switzerland}
\address[m4]{MAX IV Laboratory, Lund University, SE-221 00 Lund, Sweden}

\cortext[cor1]{Corresponding author. Telephone:  +46 46 222 9677; Fax:  +46 46 222 4709}
\fntext[fn1]{present address: Thomas Jefferson National Accelerator Facility, Newport News VA 23606, USA}
\fntext[fn2]{present address: University of Glasgow, Glasgow G12 8QQ, Scotland, UK}

\begin{abstract}
Rate-dependent effects in the electronics used to instrument the tagger focal 
plane at the MAX IV Laboratory were recently investigated using the novel 
approach of Monte Carlo simulation to allow for normalization of high-rate 
experimental data acquired with single-hit time-to-digital converters 
(TDCs). The instrumentation of the tagger focal plane has now been expanded to 
include multi-hit TDCs. The agreement between results obtained from data taken 
using single-hit and multi-hit TDCs demonstrate a thorough understanding of the 
behavior of the detector system.
\end{abstract}


\end{frontmatter}

\section{Introduction}
\label{section:introduction}

The Tagged-Photon Facility (TPF)~\cite{adler2012,tpf} at the MAX IV 
Laboratory~\cite{m4} in Lund, Sweden has been used to measure photonuclear
cross sections in many experiments. Rate-dependent deadtime and other effects
in the electronics used to instrument the tagger focal plane (FP) must be
correctly addressed in order to properly normalize the experimental data. 
These effects are particularly important because of the intermittently high 
instantaneous photon-beam flux caused by the non-uniform time structure of 
the photon beam. Limitations in the FP instrumentation electronics were also
problematic. An in-depth investigation of the rate-dependent effects at the 
TPF was recently reported~\cite{myers_nim2013} in which the novel approach 
of Monte Carlo simulation was employed. The behavior of the FP instrumentation
electronics was successfully modeled for each detected electron in 1~ns steps.
Input parameters were taken directly from the electronics setup (such as pulse 
widths) or from the data itself (such as electron rates and time structure of 
the electron beam).

The major limitation in the original electronic instrumentation system for 
the FP which led to large corrections at high rates was the use of single-hit 
time-to-digital converters (TDCs). These TDCs were used to measure the elapsed 
time between a photon-induced reaction product and the post-bremsstrahlung 
electron corresponding to the photon in question. Understanding the behavior 
of the single-hit TDCs at high rates has enabled the absolute normalization 
of data~\cite{myers_prc2013}. Recently, the instrumentation of the tagger 
FP has been upgraded to include multi-hit TDCs; that is, TDCs which are 
sensitive to more than one stop signal they receive when triggered. Such TDCs 
are superior to their single-hit predecessors as the data they provide 
eliminate the need for a large rate-dependent correction to the absolute 
experiment normalization (see below). This in turn simplifies the data 
analysis.

In this paper, we present a detailed analysis of the behavior of the tagger 
FP instrumented with multi-hit TDCs. We compare this behavior to that of
the tagger FP instrumented with single-hit TDCs and demonstrate good 
agreement. Finally, we present absolute cross-section data obtained using 
both devices and compare it to existing data to demonstrate a thorough 
understanding of the behavior of the detector system.

\section{Facility overview} 
\label{section:facility_overview}

\begin{figure}
\resizebox{1.00\textwidth}{!}{\includegraphics{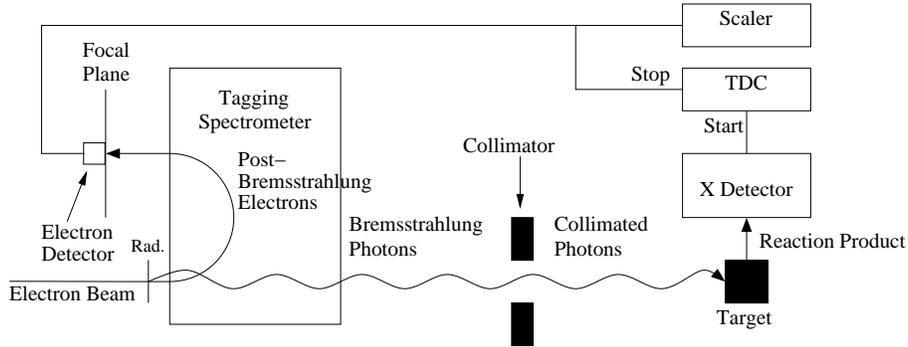}}
\caption{\label{figure:figure_01_tagging_technique}
The photon-tagging technique. Beam electrons may radiate bremsstrahlung
photons. Post-bremsstrahlung electrons are momentum analyzed using a photon 
tagger. Bremsstrahlung photons which pass through the collimator to strike the 
target and may induce photonuclear reactions. The coincidence between a 
reaction product and a post-bremsstrahlung electron is a tagged-photon event. 
Figure from~\cite{myers_nim2013}.
}
\end{figure}

At the TPF, photon taggers~\cite{vogt1993,sal1994,sal1995} and the well-known 
photon-tagging technique~\cite{adler2012,adler1997,adler1990} 
(see Fig.~\ref{figure:figure_01_tagging_technique}) are used to perform 
photonuclear investigations.  A pulse-stretched electron 
beam~\cite{lindgren2002} with an energy of up to 200~MeV is used to produce 
bremsstrahlung as it passes through a $\sim$100~$\mu$m Al radiator. The 
resulting bremsstrahlung photon beam is collimated prior to striking the 
experimental target. Post-bremsstrahlung electrons are momentum-analyzed in the
tagging spectrometer equipped with a 63-detector plastic-scintillator array 
positioned at the focal plane. A prompt coincidence between a photonuclear 
reaction product and a post-bremsstrahlung electron in the scintillator array
indicates a tagged-photon event.

The energy of the tagged photon is determined from the difference between the 
energy of the incident electron beam and the energy of the post-bremsstrahlung 
electron detected in the scintillator focal plane (FP). The measured cross
section is given by
\begin{equation}
\label{equation:ds_dw}
\frac{d\sigma}{d\Omega}=\frac{Y_{\rm coincidence}/\epsilon_{\rm detector}}{N_{\rm target} \cdot N_{\rm electrons} \cdot \epsilon_{\rm tagg} \cdot \Delta\Omega},
\end{equation}
where $Y_{\rm coincidence}$ is the number of true, prompt coincidences between 
the reaction-product detector and the FP, $\epsilon_{\rm detector}$ is the
reaction-product detector efficiency, $N_{\rm target}$ is the number of 
target nuclei per unit area, $N_{\rm electrons}$ is the number of electrons
detected in the FP array and counted in the FP scalers, $\epsilon_{\rm tagg}$
is the probability that a taggable bremsstrahlung photon passes through the 
beam-defining collimator and hits the target~\cite{adler2012}, and $\Delta\Omega$ 
is the solid angle subtended by the reaction-product detector. Both 
$Y_{\rm coincidence}$ and $N_{\rm electrons}$ must be corrected for the effects 
of deadtime in the instrumentation electronics, and the size of the corrections 
depends on the count rate.  This is complicated by the fact that the electron 
beam delivered by the accelerator has a periodic structure of varying intensity.  
As a result, the instantaneous FP rate can be almost a factor of 4 higher than 
the average FP rate (typically 3 MHz/MeV) at 20~nA, a typical average operating 
current.

\begin{figure}
\resizebox{1.00\textwidth}{!}{\includegraphics{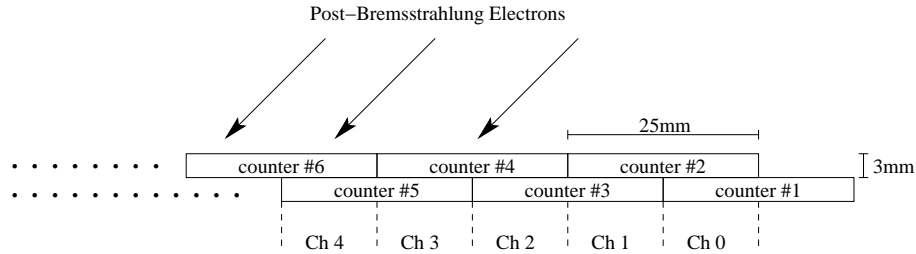}}
\caption{\label{figure:figure_02_focal_plane_hodoscope}
The FP hodoscope in 50\%-overlap configuration. A coincidence between a 
detector in the front plane and a detector in the back plane defines a tagger 
channel. There are a total of 63 detectors and thus 62 channels in the FP.
Figure from~\cite{myers_nim2013}.
}
\end{figure}

The FP hodoscope consists of two parallel rows of NE110 scintillators.  The 
front row nearest the exit window of the tagger magnet has 31 elements, while 
the back row has 32 elements 
(see Fig.~\ref{figure:figure_02_focal_plane_hodoscope}). The signals from the
detectors are passed to LRS~4413 leading-edge discriminators operated in 
burst-guard mode.  The resulting logic signals are typically set to 25 or 50~ns. 
Overlap coincidence modules designed and built at the Saskatchewan Accelerator
Laboratory (SAL) are used to identify coincidences between two physically 
overlapping detectors in the front and back rows. An output pulse is generated 
whenever the two input pulses overlap and is ended whenever one or both inputs 
are reset. An overlap of at least 3~ns is necessary to produce an output pulse.
These coincidences define FP channels and are used to stop TDCs and increment 
scalers. When a post-bremsstrahlung electron event occurs in coincidence with 
a trigger from the experiment detectors, a tagged-photon event may have occurred. 

\begin{figure}
\resizebox{1.00\textwidth}{!}{\includegraphics{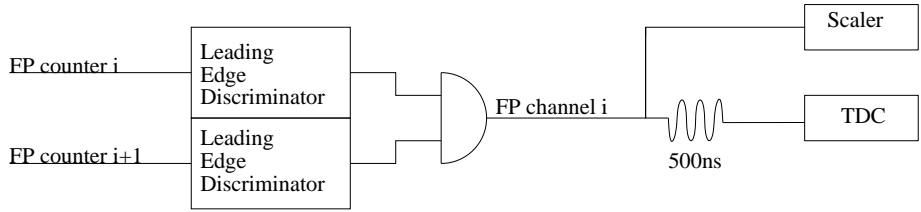}}
\caption{\label{figure:figure_03_focal_plane_electronics}
The FP electronics. A coincidence between an electron-detector signal in the
front plane of the FP and an electron-detector signal in the back plane of the 
FP defines a tagger channel. The coincidence module looking for these overlaps 
was from SAL. This signal was counted and used to stop a TDC started by the 
photonuclear reaction-product detector. 
}
\end{figure}

The device labeled TDC in Figs.~\ref{figure:figure_01_tagging_technique} and
\ref{figure:figure_03_focal_plane_electronics} represents both single-hit and 
multi-hit TDCs -- that is, both devices are used in parallel -- started by the 
same signal and stopped by the same signal(s). The single-hit TDC used to 
instrument the FP array is the CAEN~V775. The V775 is a 32 channel device with 
12~bit resolution. It is operated in common-start mode. The stop comes from the
first signal presented by a FP channel. It was experimentally determined that 
to be registered by the TDC, the stop signals corresponding to a FP channel
had to correspond to a timing overlap of at least $\sim$11~ns between the front 
row and back row signals.  The multi-hit TDC used to instrument the FP array 
is the CAEN~V1190B. The V1190B is a 64~channel device with 19~bit resolution.  
Once triggered, it uses one of the FP channel signals as the timing reference 
signal\footnote{
see http://www.caen.it/csite/CaenProd.jsp?idmod=787\&parent=11 for details}. 
The module was programmed to accept up to 4 stop signals per channel for 
each trigger.\footnote{
In order to address the rate-dependent stolen-coincidence effect (see
Sect.~\ref{subsection:stolen_coincidences}), a record of the first 2 signals 
presented by a FP channel is sufficient. We record the first 4 signals in order
to be able to better confirm that our instrumentation electronics are behaving 
as expected.}. 
It was experimentally determined that these stop signals also had to be at 
least $\sim$11~ns in width. The device labeled scaler in 
Figs.~\ref{figure:figure_01_tagging_technique} and 
\ref{figure:figure_03_focal_plane_electronics} is a CAEN~V830 scaler. The
V830 is a 32~channel latching device with a 250~MHz counting capability.  It
was experimentally determined to register pulses a short as $\sim$3~ns in 
width, the limit of our FP-trigger setup.

Two advantages to requiring a coincidence between the front and back rows of
electron detectors in the FP array are that registration of the background in 
the experimental hall is greatly suppressed, and that the photon-energy 
resolution may be easily increased simply by offsetting the two scintillator 
planes.

\section{Rate-dependent effects}
\label{section:rate_dependent_effects}

As the electron beam has a varying periodic intensity, high instantaneous 
post-bremsstrahlung electron event rates can occur. The resulting rate-dependent 
effects may result in significant losses in the number of events registered by 
the FP array instrumentation. Unless taken carefully into consideration, these 
rate-dependent effects prevent the absolute normalization of the experimental 
data. Rate-dependent effects include ghost events, missed stops, and stolen 
coincidences (see below). Table~\ref{table:table_01_corrections_summary} 
summarizes typical values for rate-dependent corrections to the number of 
post-bremsstrahlung electrons detected by the FP array for both the single-hit 
and multi-hit TDCs.

\begin{table}
\caption{\label{table:table_01_corrections_summary}
A summary of corrections to the number of tagged events registered by the FP 
array required for the absolute normalization of experimental data. Note that 
the stolen-coincidences correction is unnecessary when multi-hit TDCs are used, 
a distinct advantage. See text for details.
}
\begin{center}
\begin{tabular}{ccc} \hline \hline
Correction          & single-hit TDCs & multi-hit TDCs \\
\hline
Ghost events        &             1\% &            1\% \\
Missed stops        &             3\% &            3\% \\
Stolen coincidences &            40\% &            N/A \\
\hline \hline
\end{tabular}
\end{center}
\end{table}

\subsection{Ghost events}
\label{subsection:ghost_events}

\begin{figure}
\resizebox{1.00\textwidth}{!}{\includegraphics{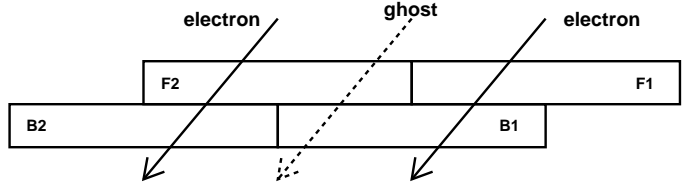}}
\caption{\label{figure:figure_04_ghosts}
Creation of a ghost event. Real post-bremsstrahlung electrons (solid arrows) 
in next-to-neighboring FP channels arrive at almost the same time.  This 
creates the illusion -- or ghost -- of an electron (dashed arrows) in the 
counters that constitute the intermediate FP channel. Figure 
from~\cite{myers_nim2013}.
}
\end{figure}

A major disadvantage of requiring a coincidence between the front and back rows
of electron detectors in the FP array is the creation of ghost events at high 
rates. The ghost events result from the instrumentation of the FP array.  The 
scenario leading to a ghost event is illustrated in 
Fig.~\ref{figure:figure_04_ghosts}. Two different post-bresmsstrahlung 
electrons strike next-to-neighboring channels (counters F1~$\cdot$~B1 and 
F2~$\cdot$~B2) at nearly the same time which creates the illusion of an 
electron in the channel in between (counters F1~$\cdot$~B2) -- the ghost event.
The rate of the accidental coincidences that result in ghost events depends on 
the post-bremsstrahlung electron rate, the widths of the FP discriminator output 
pulses, and the resolving time of the overlap coincidence modules. Because 
these ghosts are formed in the FP electronics, they are registered as 
coincidences in both the FP scalers and the FP TDC modules, resulting in a 
partial but not complete cancellation of the effect. As the rate of ghost events 
is purely a function of post-bresmsstrahlung electron rate and FP geometry, they 
affect both single-hit and multi-hit TDC data equally. They are best addressed 
using the simulation approach detailed in Ref.~\cite{myers_nim2013}.

\subsection{Missed stops}
\label{subsection:missed_stops}

A missed stop occurs when the FP scalers register a recoil-electron event
while the FP TDCs miss it. The primary origin of missed stops lies in the 
different minimum pulse width for registration in the FP scalers ($\sim$3~ns) 
and TDCs ($\sim$11~ns).  This deadtime effect is best addressed using the 
simulation approach detailed in Ref.~\cite{myers_nim2013}.

\subsection{Stolen coincidences}
\label{subsection:stolen_coincidences}

\begin{figure}
\resizebox{1.00\textwidth}{!}{\includegraphics{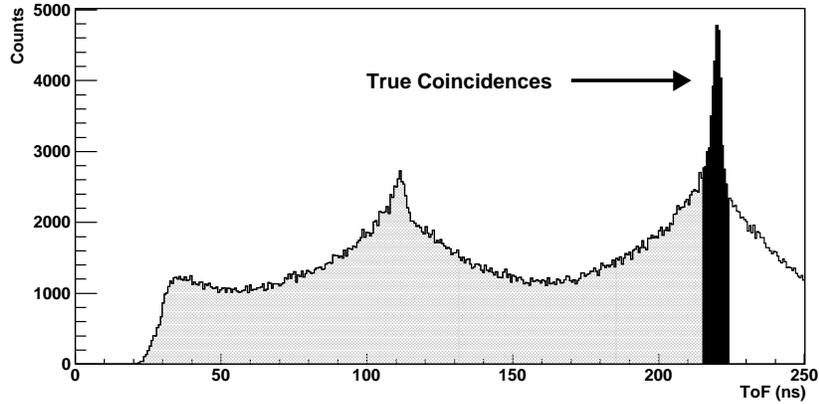}}
\caption{\label{figure:figure_05_stolen_coincidences}
The stolen-coincidence effect in a single-hit TDC spectrum acquired at a high 
post-bremsstrahlung electron rate. The black peak at channel 225 represents 
true coincidences between the reaction-product detector and the FP array, and
is the earliest possible time that a true coincidence may be registered. Events 
in the lightly shaded region correspond to accidental post-bremsstrahlung 
electrons that stop the FP TDCs before this earliest possible point in time. 
The coincidence is thus mis-timed and the true coincidence event is stolen when
a single-hit TDC is used. Note that the ``peak'' at channel 110 is an artifact 
of the extracted electron beam. Figure from~\cite{myers_nim2013}.
}
\end{figure}

When single-hit TDCs are employed, due to the fact that the TDC only registers
the first signal presented to it subsequent to the start, an accidental
post-bremsstrahlung electron may be detected in the FP channel before the 
actual post-bremsstrahlung electron that corresponds to the tagged photon. The 
result is that the single-hit TDC is stopped too early, leading to a 
well-studied phenomenon known as stolen coincidences -- see 
Fig.~\ref{figure:figure_05_stolen_coincidences}. Well-known 
methods~\cite{owens1990,hoorebeke1992} exist for determining the 
stolen-coincidence correction. It may also be efficiently addressed using the 
simulation approach detailed in Ref.~\cite{myers_nim2013} or greatly reduced
by implementing multi-hit TDCs.

As previously mentioned, the multi-hit TDC with which the FP has been upgraded
has been programmed to accept the first 4 stop signals presented to it in 
conjunction with the trigger and the programmable acceptance window. In this 
manner, the stolen-coincidence effect is greatly reduced as up to 3 accidental 
post-bresmsstrahlung electrons can be registered in the multi-hit TDC before the 
prompt electron without stealing it. Thus, a large correction to the number of 
tagged events registered by the FP array is avoided, increasing the precision 
of the overall normalization.\\

\begin{figure}
\resizebox{1.00\textwidth}{!}{\includegraphics{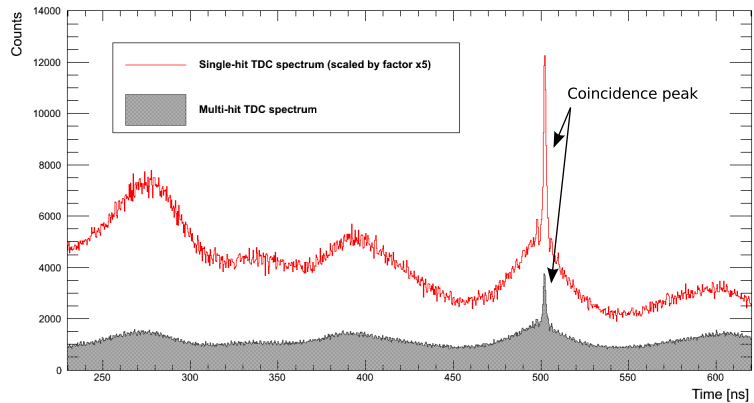}}
\caption{\label{figure:figure_06_sh_mh_comparison}
A comparison between single-hit (red, unshaded) and multi-hit (gray, shaded) FP
TDC spectra obtained simultaneously. Note the $\times$5 scaling of the 
single-hit TDC spectrum. The time structure in the spectra is due to the method 
by which the beam is generated. The peak representing coincidences between the 
FP and the reaction-product detector is clearly evident at $\sim$500~ns. The 
slope of the background in the unshaded single-hit TDC spectrum is proportional 
to the post-bremsstrahlung electron rate and also to the magnitude of the 
stolen-coincidence correction. As expected, no slope is evident in the background 
in the shaded multi-hit TDC spectrum as coincidences may not be stolen. (For 
interpretation of the references to color in this figure caption, the reader is 
referred to the web version of this article.)
}
\end{figure}

The causes of ghost events, missed stops, stolen coincidences, and deadtime in
the FP scalers and TDCs are all included in the FP simulation~\cite{myers_nim2013}
which can therefore be used to determine suitable correction factors.

\section{Results}
\label{section:results}

\begin{figure}
\resizebox{1.00\textwidth}{!}{\includegraphics{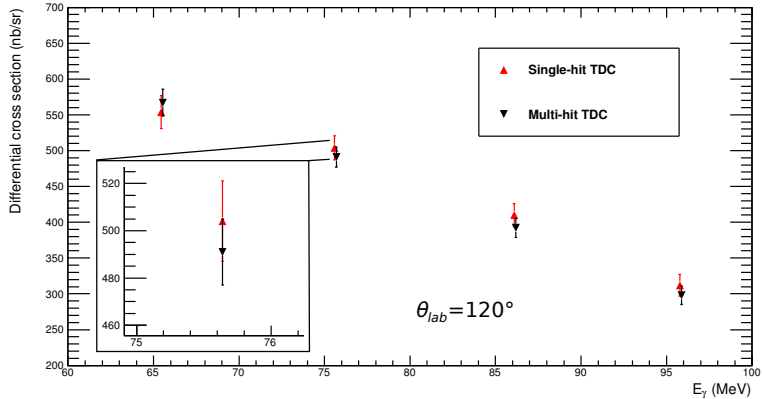}}
\caption{\label{figure:figure_07_TDC_cross_section}
Absolute differential cross-section data for elastic photon scattering from 
$^{12}$C at a lab angle of 120$^{\circ}$ obtained simultaneously using both 
single-hit (upright red triangles) and multi-hit (inverted black triangles) 
TDCs. The single-hit TDC data have been corrected according to 
Ref.~\cite{myers_nim2013}. Statistical uncertainties only are shown. See text 
for details. (For interpretation of the references to color in this
figure caption, the reader is referred to the web version of this article.)
}
\end{figure}

Figure~\ref{figure:figure_07_TDC_cross_section} presents a comparison
between the absolute differential cross section for elastic photon scattering
from $^{12}$C at a lab angle of 120$^{\circ}$ recently obtained at the
TPF at the MAX IV Laboratory using both single-hit (upright red triangles)
and multi-hit TDCs (inverted black triangles) simultaneously. The single-hit
TDC data have been corrected for stolen coincidences according to 
Ref.~\cite{myers_nim2013}. Error bars reflect statistical uncertainties only. 
The agreement between the two data sets is satisfactory, thereby confirming our
understanding of the stolen-coincidence correction to the data.

\begin{figure}
\resizebox{1.00\textwidth}{!}{\includegraphics{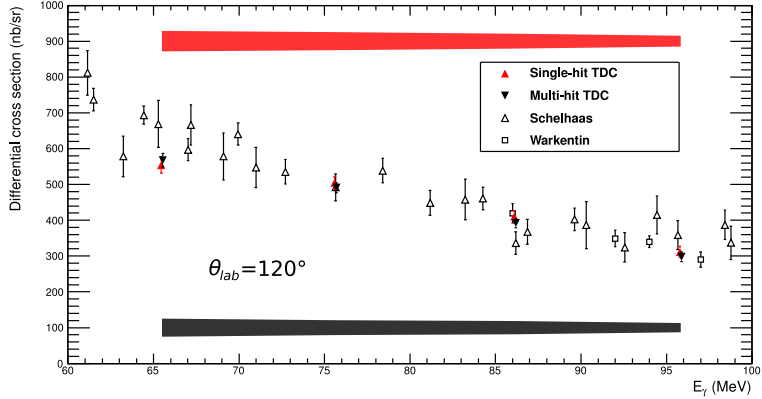}}
\caption{\label{figure:figure_08_results_cross_section}
Absolute differential cross-section data for elastic photon scattering from 
$^{12}$C at a lab angle of 120$^{\circ}$ obtained simultaneously using both
single-hit (upright red triangles) and multi-hit (inverted black triangles) 
TDCs compared to published data. The single-hit TDC data have been corrected 
according to Ref.~\cite{myers_nim2013}.  Statistical uncertainties only are 
shown. Systematic uncertainty bands for the present measurement are presented 
at the top (corrected single-hit TDC data) and bottom (multi-hit TDC data) of 
the figure.  See text for details. (For interpretation of the references to 
color in this figure caption, the reader is referred to the web version of this
article.)
}

\end{figure}

Figure.~\ref{figure:figure_08_results_cross_section} presents a comparison 
between the absolute differential cross section for elastic photon scattering 
from $^{12}$C at a lab angle of 120$^{\circ}$ recently obtained at the TPF at 
the MAX IV Laboratory and existing data published by 
Schelhaas~{\it et al.}~\cite{schelhaas1990} (open upright triangles) and 
Warkentin~{\it et al.}~\cite{warkentin2001} (open squares). Error bars reflect 
statistical uncertainties only. Systematic uncertainty bands for the present
measurement are presented at the top (red, single-hit TDC data) and bottom 
(gray, multi-hit TDC data) of the figure. The agreement between the data sets 
is very good thereby confirming our understanding of the absolute normalization.

\section{Summary}
\label{section:summary}

In this paper, the Monte Carlo simulation~\cite{myers_nim2013} of the MAX IV
tagger focal-plane electronics has been tested by comparing results obtained
using both single- and multi-hit TDCs. Good agreement between these data sets 
has been demonstrated, and the measured absolute cross sections also agree 
with previous experiments.  We conclude that the behavior of the detector 
system is thoroughly understood and that the Monte Carlo simulation 
incorporates it correctly.

\section*{Acknowledgements}
\label{acknowledgements}

This project was supported by the US National Science Foundation Grant 
No.\ 0855569, the UK Science and Technology Facilities Council, as well as 
The Swedish Research Council, the Crafoord Foundation, and the Royal 
Physiographic Society in Lund. The authors gratefully acknowledge the Data 
Management and Software Centre, a Danish Contribution to the European 
Spallation Source ESS AB, for generously providing access to their
computations cluster. We also thank J.C. McGeorge for useful discussions.

\bibliographystyle{elsarticle-num}

\end{document}